\documentclass[twocolumn,showpacs,preprintnumbers,amsmath,amssymb]{revtex4}

\usepackage{graphicx}
\usepackage{dcolumn}
\usepackage{bm}

\def\spose#1{\hbox to 0pt{#1\hss}}
\def\approxlt{\mathrel{\spose{\lower 3pt\hbox{$\sim$}}
        \raise 2.0pt\hbox{$<$}}}
\def\approxgt{\mathrel{\spose{\lower 3pt\hbox{$\sim$}}
        \raise 2.0pt\hbox{$>$}}}

\def\multleft#1{\hbox to size{\vbox {\halign {\lft{##}\cr #1}}\hfill}\par}
\def\multright#1{\hbox to size{\vbox {\halign {\rt{##}\cr #1}}\hfill}\par}

\def\boxit#1{\vbox{\hrule\hbox{\vrule\kern3pt\vbox{\kern3pt
          #1 \kern3pt}\kern3pt\vrule}\hrule}}

\begin{document}

\preprint{APS/123-QED}

\title{Nonlinear Development of Streaming Instabilities In Strongly Magnetized Plasmas }

\author{H. Che}
\author{ J. F. Drake}
\author{M. Swisdak}
\affiliation{ IREAP, Department of Physics, University of Maryland, College Park, MD, 20742, USA}
\author{P. H. Yoon}
\affiliation{IPST, University of Maryland, College Park, MD, 20742, USA}

\date{\today}

\begin{abstract}
The nonlinear development of streaming instabilities in the current layers formed during magnetic reconnection with a guide field is explored. Theory and 3-D particle-in-cell simulations reveal two distinct phases. First, the parallel Buneman instability grows and traps low velocity electrons. The remaining electrons then drive two forms of turbulence: the parallel electron-electron two-stream instability and the nearly-perpendicular lower hybrid instability. The high velocity electrons resonate with the turbulence and transfer momentum to the ions and low velocity electrons. 
\end{abstract}

\pacs{52.35.Vd, 52.35 Py, 52.35 Qz}
\maketitle
Magnetic reconnection in collisionless plasmas drives strong currents near the x-line and separatrices. Satellite observations in the Earth's magnetosphere indicate that these current layers are turbulent. Electron holes, which are localized, positive-potential structures, have been linked to magnetic reconnection in the magnetotail \cite{farrell02grl,cattell05jgr}, the magnetopause\cite{matsumoto03grl}, and the laboratory\cite{fox08prl}. Lower-hybrid waves and other plasma waves appear in conjunction with the electron holes in the magnetotail event. The dissipation associated with these turbulent wavefields may facilitate the breaking of magnetic field lines during collisionless magnetic reconnection \cite{galeev84book2}.
 
During reconnection with a guide magnetic field perpendicular to the plane of reconnection, the width of the electron current layers are of the order of the electron Larmor radius $\rho_e = v_{te}/\Omega_{e}$, where $v_{te}$ is the electron thermal velocity and $\Omega_e$ is the electron cyclotron frequency\cite{hesse04pop,rogers07pop}. The resulting streaming velocity $v_{d}$ is given by $v_d/v_{te} \propto \bigtriangleup B/(B\beta_e)$, where  $\bigtriangleup B$ is the amplitude of the reconnection magnetic field and $\beta_e = 8\pi nT_e/B^2$ is the ratio of the electron and magnetic field pressures. The Buneman instability  is driven unstable by electron-ion streaming when $v_d/v_{te} >1$ \cite{buneman58prl,papa76pop, drake03sci}. If the instability is strong enough,  the wave potential forms intense localized electric field structures called electron holes. The evolution of streaming instabilities has been previously investigated analytically and numerically \cite{ boris70prl,davidson75pof,oppenheim01grl,singh01jgra,singh01jgrb,drake03sci,omura03jgr,mcmillan06pop, mcmillan07pop,goldman08grl}. Lower hybrid turbulence is found to emerge in simulations after saturation of the parallel Buneman instability \cite{drake03sci,mcmillan06pop, mcmillan07pop}. Whether the lower hybrid turbulence is a linearly unstable mode or nonlinearly driven by electron holes is unclear \cite{mcmillan06pop, mcmillan07pop}. An important question is how the Buneman instability, which has a very low phase speed, can stop the runaway of high velocity electrons. The resolution of this puzzle is essential to understand how the streaming kinetic energy of electrons is transformed into electron thermal energy, what mechanism sustains electron holes  after saturation of the Buneman instability, and ultimately what role  turbulence plays in magnetic reconnection. 

In this letter, we investigate these questions using both numerical and analytic methods. Two distinct phases of the dynamics are discovered. In the first the parallel Buneman instability grows and traps the lower velocity streaming electrons. In the second the high velocity electrons drive two distinct forms of turbulence: the parallel electron-electron two-stream instability and the lower hybrid instability. The high parallel phase speed of these waves allows them to  resonate with high energy electrons and transfer momentum to  the ions and low velocity electrons.
 
We carry out 3D PIC simulations with strong electron drifts in an inhomogenous plasma with a strong guide field. The initial state mimics the current sheet near an x-line at late time in 2D simulations of reconnection \cite{drake03sci}. Since we apply no external perturbations to initiate reconnection, and the time scale for reconnection to occur naturally in the simulation is much longer than the growth time scale of other instabilities, reconnection does not develop. The initial conditions for our simulation are based on \cite{drake03sci}. We specify the reconnecting magnetic  field to be $B_x/B_0=\tanh[(y-L_y/2)/w_0]$, where $B_0$ is the asymptotic amplitude of $B_x$ outside of the current layer, and $w_0$ and $L_y$ are the half-width of the initial current sheet and the box size in the $y$ direction, respectively. The guide field $B_z^2 = B^2-B_x^2$ is  chosen so that the total field $B$ is  constant. In our simulation, $B$ is taken as $26^{1/2} B_0$. Thus, we are in the limit of a strong guide field such that $\Omega_e=2.5 \omega_{pe}$, where $\omega_{pe}$ is the electron plasma frequency. The initial temperature is $T_e =T_i =0.04 m_i c_A^2$, the ion to electron mass ratio is $100$, the speed of light $c$ is $20 c_A $, and $c_A= B_0/(4 \pi n_0 m_i)^{1/2}$ is the Alfv\'en speed.  The simulation domain has dimensions $L_x=L_y=d_i= c/\omega_{pi}$, and $L_z=8 d_i$, where $\omega_{pi}$ is the ion plasma frequency. The initial electron drift along $\hat{z}$ is $10 c_A$, close to that of the current layer around the x-line of 2D reconnection simulations reported earlier \cite{drake03sci}. The initial ion drift is $-0.9 c_A$. The drift speed in these simulations is around three times the electron thermal speed $v_{te} \sim 3 c_A$, above the threshold to trigger the Buneman instability.
\begin{figure}
\includegraphics[scale=0.5, trim=70 200 0 20,clip]{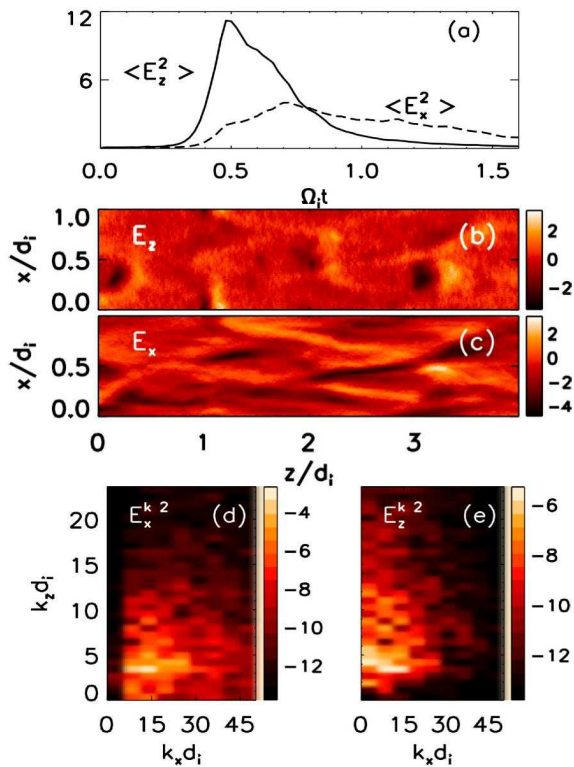} 
\caption{In (a): the time evolution of $\langle E_x^2\rangle$ (dashed line) and $\langle E_z^2\rangle$ (solid line). In (b) and (c): the spatial structure of $E_z$  and $E_x$  at $\Omega_i t =1.2$. In (d) and (e): the corresponding 2D power spectra $|E_x(k_x,k_z)|^2$ and $|E_z(k_x,k_z)|^2$. The power spectra are shown on a logarithmic scale. }
\label{simspec}
\end{figure}

In our simulation the intense electron stream drives a strong Buneman instability within a cyclotron period. In Fig.~\ref{simspec}~(a), we show the time development of $\langle E_x^2\rangle$ and $\langle E_z^2\rangle$, where $\langle\rangle$ denotes an average over the midplane of the current sheet. As the instability  develops the Buneman waves clump to form localized electron holes.  At the same time, the electric field also becomes stretched and forms long oblique stripes in the $x-z$ plane. This behavior is reflected in the delayed growth of $\langle E_x^2\rangle$ in Fig. \ref{simspec} (a).  In Fig.~\ref{simspec}~(b) and (c), we show snapshots  of the components  $E_x$ and $E_z$ of the electric field at the midplane of the current layer at $\Omega_i t=1.2$. The two distinct structures are evident. As in earlier simulations \cite{drake03sci}, the electron holes are round and therefore in $k$-space have $k_x \sim k_z$. The wavevectors of the stripes are oriented $\sim 80^{\circ}$ relative to the magnetic field. The power spectra of $E_x$ and $E_z$ in Fig.~\ref{simspec}~(d,e) reveal that the power spectrum of $E_x$ is peaked around $(k_x, k_z)\sim (10, 5)$ with a width $\delta k_x \sim 10$, which corresponds to the spatial scale of the stripes in Fig.~\ref{simspec}~(c). The peak in the power spectrum of $E_z$ is less well defined. One peak corresponds to the stripes, and  the other is broadly centered around $(k_x, k_z)\sim (0, 10)$ with a width $\delta k_x \sim 10$. This corresponds to the spatial scale of the electron holes. The spectra suggest that two modes have developed at late time.
\begin{figure}
\includegraphics[scale=0.36,trim=10 140 0 30,clip]{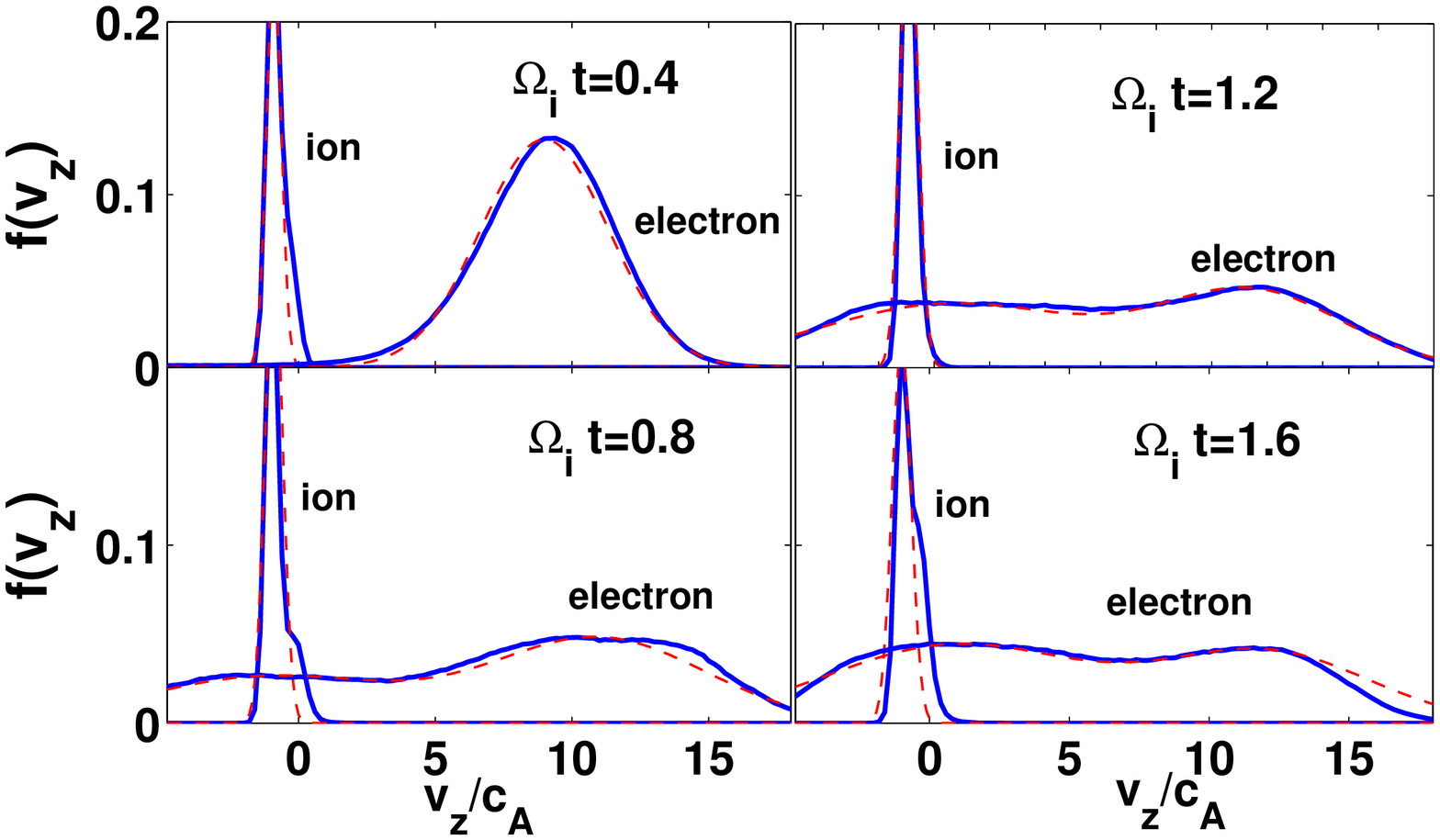} 
\caption{ Distribution functions $f(v_z)$ of ions and electrons from the PIC simulation at $\Omega_i t = 0.4, 0.8, 1.2, 1.6$ (solid blue) with the ion distribution function reduced by a factor of $4$. The theoretical fitting with double drifting Maxwellians (dashed red).}
\label{dist}
\end{figure}

\begin{figure}
\includegraphics[scale=0.46, trim=50 470 0 50 ]{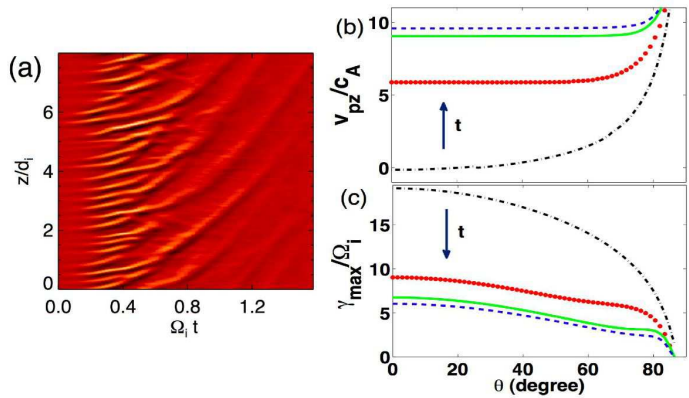} 
\caption{In (a) cuts of $E_z$ versus $z$ and time from the simulation. The slope of the curves  is the phase speed $v_{pz}$ in the simulation rest frame. In (b) and (c) solutions of the dispersion relation using the fittings of distribution functions shown in Fig. \ref{dist} at $\Omega_i t= 0.4, 0.8,1.2, 1.6$. In (b), the parallel phase speed versus the angle $\theta$ between the wavevector $\mathbf{k}$ and the magnetic field. In (c), the corresponding maximum growth rates $\gamma_{max}$ versus $\theta$. }
\label{phavel}
\end{figure}

In the cold plasma limit, the phase velocity of the Buneman instability is  $(m_e/(2 m_i))^{1/3}|v_{ez}|/2 + v_{di}\sim -0.25 c_A $ \cite{galeev84book1}. The electrons with velocity around this phase velocity can be trapped. Evidence for trapping can be seen from the change in the electron distribution functions  between $\Omega_i t= 0.4$ and $0.8$ in Fig.~\ref{dist}. The distribution function at $\Omega_i t=0.8$ has a broad peak centered around the expected Buneman phase speed. Some of the high velocity electrons actually increase their velocity $v_z$ early in the simulation because they are accelerated locally by the induced electric field $E_z$ that maintains the net current in the layer.  At times $\Omega_i t= 1.2$ and $1.6$ the electrons at higher velocity are dragged to lower velocity. This behavior can not be explained by wave-particle interactions with the Buneman instability. In our simulation, the average drift velocity of electrons decreases by a factor of two from the initial value $\sim 10 c_A$ to $\sim 5 c_A$ at $\Omega_i t=1.6$. To understand the late time behavior of the electrons, we investigate the phase speed of $E_z$ by stacking cuts of $E_z(z)$ at time intervals $\Omega_i t=0.02$. The image shown in Fig. \ref{phavel} (a) traces the motion  of the peaks and valleys of the waves. The slopes of the curves formed by the peaks or valleys  show the evolution of the phase speed in the $z$ direction. The slopes of these curves increase with time, indicating that the parallel phase speed $v_{pz}$ of the waves increases. The $v_{pz}$ increases from a very small value at $\Omega_i t = 0.4$ when the instability onsets to $6 c_A$ at $\Omega_i t = 0.8$, and nearly $10 c_A$ at the end of the simulation. The phase speed of $E_x$  (not shown) along $z$ behaves similarly. Thus, we suggest that the increasing phase speed of the turbulence at late time allows the electrons with high velocity to be dragged to lower velocity  through wave-particle interactions. The early trapping of electrons by the Buneman instability saturates and  allows new instabilities grow.

An interesting feature of the waves in our simulation is the growth of two distinct classes of modes parallel and nearly-perpendicular to the magnetic field. We explore the nature of the late-time instabilities by investigating the unstable wave spectra with particle distributions that match those measured in the simulations. We choose double drifting Maxwellians for electrons, and a single drifting Maxwellian for ions. 

\begin{figure}
\includegraphics[scale=0.35,trim=50 45 0 55 ,clip]{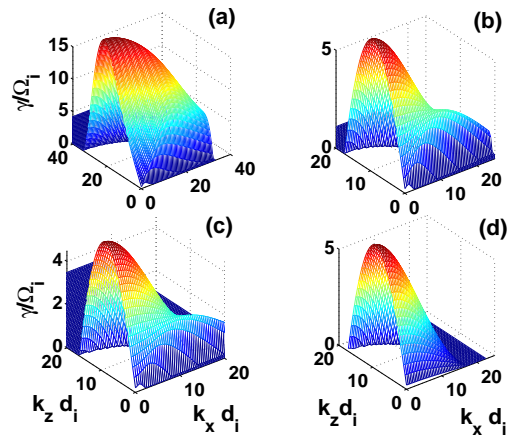} 
\caption{The 2D theoretical spectra from kinetic theory using the distribution function fittings shown in Fig.~\ref{dist}. In (a), the spectrum at $\Omega_i t =0.4$ when the electron distribution is still a single drifting Maxwellian. In (b) and (c), the spectra at $\Omega_i t =1.2$ and $1.6$ using the distributions of both electrons and ions. In (d), the spectrum again at $\Omega_i t=1.2$ but neglecting the ions.}
\label{matspec}
\end{figure}

The local dispersion relation is, for waves with $\Omega_{i} \ll \omega \ll \Omega_e$, (Che et al. 2009, {\it in prep}):
\begin{multline}
1+\frac{2\omega_{pi}^2}{k^2 v^2_{ti}}[1+\zeta_i Z(\zeta_i)] 
+ \frac{2 \delta\omega^2_{pe}}{k^2 v^2_{te1}}[1+I_0(\lambda) e^{-\lambda} \zeta_{e1} Z(\zeta_{e1})]\\
+ \frac{2 (1-\delta)\omega^2_{pe}}{k^2 v^2_{te2}}[1+I_0(\lambda) e^{-\lambda} \zeta_{e2} Z(\zeta_{e2})]=0,
\label{bdf}
\end{multline}
where  $\zeta_i=(\omega-k_{z} v_{di})/k v_{ti}$, $\zeta_{e1} = (\omega-k_{z} v_{de1})/k_{z} v_{z te1}$, $\zeta_{e2} = (\omega-k_{z} v_{de 2})/k_{z} v_{z te2}$, $\lambda=k^2_x v^2_{x te}/2\Omega^2_e$, $\delta$ is the weight of the high velocity drifting Maxwellian, $Z$ is the plasma dispersion function and $I_0$ is the modified Bessel function of the first kind with order zero. The thermal velocity of species $j$ is defined by  $v^2_{tj}=2T_{tj}/m_j$ and drift speed by $v_{dj}$, which is parallel to the magnetic field ($z$ direction). The ions are taken to be isotropic. It should be noted that our simulation is highly non-linear  while our model is based on linear theory. This approach is reasonable, however, because we can fit the real time distribution functions from the simulation for input into the dispersion relation in Eq.~(\ref{bdf}) to obtain the wave modes. In Fig.~\ref{dist} we show that the real time distribution functions of the electrons (blue lines) are well modelled by a double drifting Maxwellian (dashed red lines). The parameters of the fittings are listed in the Table~\ref{bunpara}. In Fig.~\ref{matspec}~(a,b,c) we show the growth rate of unstable modes, obtained from Eq.~(\ref{bdf}), in the $(k_x, k_z)$ plane at $\Omega_i t=0.4,1.2,1.6$.  
\begin{table}[h]
\caption[Short title]{Parameters of Model Dist. Funs.}
\begin{center}
\begin{tabular}{|c|c|c|c|c|c|c|c|}
\hline  & $v_{zte1}$ & $v_{zte2}$ & $v_{de1}$ & $v_{de2}$ & $v_{ti}$  & $v_{di}$  & $\delta$ \\ 
\hline $\Omega_i t$=0.4 & 3.3 &  & 9 &  & 0.38 & -0.9 &  1\\ 
\hline $\Omega_i t$=0.8 & 5.5 & 6.6 & 11 & -1 & 0.45 & -0.9 & 0.6 \\ 
\hline $\Omega_i t$=1.2 & 4.6 & 7 & 11.5 & 0.7 & 0.45 & -0.9 & 0.43 \\ 
\hline $\Omega_i t$=1.6 & 5.2 & 6.6 & 12.1 & 1 & 0.48 & -0.9 & 0.41 \\ 
\hline
\end{tabular} 
\end{center}
\label{bunpara}
\end{table}

From Fig.~\ref{dist} we see that at $\Omega_i t=0.4$, around the onset of the Buneman instability, the electron distribution function is well approximated by a single drifting Maxwellian. The unstable modes shown in the spectrum in Fig.~\ref{matspec}~(a) are characteristic of the Buneman instability. The dominant mode has $\mathbf{k}$ parallel to $\mathbf{B}$ with $k_z \sim 20$ and with a large growth rate  $\gamma \sim 15 \Omega_i$. At $\Omega_i t=1.2$, the electron distribution function in Fig. ~\ref{dist}~(c) has a double peak. Interestingly,  two distinct modes appear  in Fig.~\ref{matspec}~(b). The stronger mode is peaked at $k_x \sim 0$ and $k_z \sim 10$. The weaker mode is peaked at $k_x \sim 20$ and $k_z \sim 5$. These unstable spectra are consistent with the spectra of $E_z$ and $E_x$ shown in Fig.~\ref{simspec}~(d,e). At late time, $\Omega_i t=1.6$, both modes become weaker, as shown in Fig.~\ref{matspec}~(c). 

Do these two modes  belong to the general class of Buneman instabilities or do they represent  the emergence of a new class of instabilities? We have shown that the electrons can be modeled with double drifting Maxwellians, suggesting that an electron-electron two-stream instability might develop. To investigate this possibility, we solve the dispersion relation in Eq.~(\ref{bdf}) by removing the contribution from the ions. The unstable modes for the resulting two-stream instability are shown in Fig.~\ref{matspec}~(d). Indeed, the growth rate of the two-stream instability corresponds to the stronger mode shown in Fig.~\ref{matspec}~(b). The corresponding frequency is about $0.7 \omega_{pe}$ in the ion rest frame. The weaker mode is not present in Fig.~\ref{matspec}~(d) and must involve the ions. Its wavevector is nearly perpendicular to the direction of the magnetic field. The frequency for this mode is around $16 \Omega_i $, which is close to the lower-hybrid frequency in the cold plasma limit, $\omega_{lh}=\omega_{pi}/(1+\omega^2_{pe}/\Omega_e^2)^{1/2} + k_z v_{di} \sim 15 \Omega_i$. We thus conclude that the weaker mode is in fact a current driven lower hybrid instability, which was discussed earlier by McMillan and Cairns \cite{mcmillan06pop, mcmillan07pop}. The Buneman instability in our simulation has evolved into a dual state of electron two-stream and lower hybrid instabilities.

The drag on the high velocity electrons requires wave-particle interactions with waves of high phase speed. In Fig.~\ref{phavel} we show the phase speed of unstable waves using the dispersion relation in Eq.~(\ref{bdf}) and the model distribution functions. At a given angle $\theta$ between the wavevector $\mathbf{k}$ and  magnetic field $\mathbf{B}$, the maximum growth rate is calculated with respect to the magnitude of $k$ and is shown in Fig.~\ref{phavel}~(c). The resulting parallel phase speed $v_{pz}$ versus $\theta$ is shown in Fig.~\ref{phavel}~(b). The four lines are the phase speed calculated at times $\Omega_i t=0.4,0.8,1.2,1.6$ (black dashed, red dotted, green solid, blue dashed, respectively). The phase speed increases with time, especially at small values of $\theta$, transitioning from the Buneman to the electron two-stream instability. At the times shown the phase speeds at small $\theta$ are around $0,6,9$ and $10 c_A$, consistent with the simulation data in Fig.~\ref{phavel}~(a). The phase speeds at large $\theta$ are much larger than at small $\theta$, even at $\Omega_i t=0.4$, but as shown in Fig.~\ref{phavel}~(c) the corresponding maximum growth rate at this early time is small compared with the Buneman growth rate. The growth rates at large angle become comparable to the rates at small angle at late time, which is consistent with the development of transverse modes in the simulation at late time.

To summarize, we have shown that in strongly magnetized plasmas, the parallel Buneman instability  evolves into two distinct instabilities: the parallel two-stream instability and the nearly-perpendicular lower hybrid instability.  The two-stream instability continuously sustains the electron holes that first formed due to the Buneman instability while the lower hybrid instability drives turbulence in the perpendicular direction. The high phase speed of the waves at late time couples the highest velocity streaming electrons to the ions and low velocity electrons. During reconnection the electron two-stream and lower hybrid instabilities might prevent electron runaway, facilitating the breaking of the frozen-in condition required for fast reconnection. The observations of electron holes simultaneously with perpendicularly polarized, lower-hybrid waves and parallel-polarized plasma waves with frequency close to $\omega_{pe}$ during reconnection in the magnetotail \cite{cattell05jgr} are consistent with the present model. We note, however, the observations correspond to $\omega_{pe}/\Omega_e \sim 4$ while in our simulations $\omega_{pe} < \Omega_e$. In similar simulations with $\Omega_{e}=\omega_{pe}$ two stream and lower hybrid turbulence are also excited. An interesting possibility is that in the limit $\omega_{pe} \gg \Omega_e$, the Kadomtsev-Pogutse instability might play a role in scattering the high velocity electrons \cite{liu77prl}.
 
This work was supported in part by NSF ATM0613782, and NASA NNX08AV87G and NWG06GH23G. PHY acknowledges NSF grant ATM0836364. The simulations were carried out at the National Energy Research Scientific  Computing Center.

\newpage 
\bibliography{plasma}
\end{document}